\documentclass[useAMS,usegraphicx]{mn2e}
\usepackage{graphicx}

\usepackage{amsmath} 
\usepackage{amssymb}
\usepackage{amsfonts} 
\usepackage{amscd} 
\usepackage{enumerate}
\usepackage{mathrsfs} 

\def\ltsima{$\; \buildrel < \over \sim \;$}
\def\simlt{\lower.5ex\hbox{\ltsima}}

\title{On the equilibrium of rotating filaments}

\author[Recchi et al.]{S.
  Recchi$^{1}$\thanks{simone.recchi@univie.ac.at}, A. Hacar$^{1}$
  \thanks{alvaro.hacar@univie.ac.at} and A. Palestini$^{2}$
  \thanks{arsen.palestini@uniroma1.it}\\
  $^{1}$ Department of Astrophysics, Vienna University,
  T\"urkenschanzstrasse 17, A-1180, Vienna, Austria \\
  $^{2}$ MEMOTEF, Sapienza University of Rome Via del Castro
  Laurenziano 9, 00161 Rome, Italy}

\date{Received; accepted}

\begin{document}
\maketitle


\begin{abstract}
  The physical properties of the so-called Ostriker isothermal,
  non-rotating filament have been classically used as benchmark to
  interpret the stability of the filaments observed in nearby clouds.
  However, such static picture seems to contrast with the more
  dynamical state observed in different filaments. In order to explore
  the physical conditions of filaments under realistic conditions, in
  this work we theoretically investigate how the equilibrium structure
  of a filament changes in a rotating configuration.  To do so, we
  solve the hydrostatic equilibrium equation assuming both uniform and
  differential rotations independently.  We obtain a new set of
  equilibrium solutions for rotating and pressure truncated filaments.
  These new equilibrium solutions are found to present both radial and
  projected column density profiles shallower than their Ostriker-like
  counterparts. Moreover, and for rotational periods similar to those
  found in the observations, the centrifugal forces present in these
  filaments are also able to sustain large amounts of mass (larger
  than the mass attained by the Ostriker filament) without being
  necessary unstable.  Our results indicate that further analysis on
  the physical state of star-forming filaments should take into
  account rotational effects as stabilizing agents against gravity
\end{abstract}


\maketitle


\section{Introduction}\label{sec:intro}

Although the observations of filaments within molecular clouds have
been reported since decades (e.g. Schneider \& Elmegreen 1979), only
recently their presence has been recognized as a unique characteristic
of the star-formation process.  The latest Herschel results have
revealed the direct connection between the filaments, dense cores and
stars in all kinds of environments along the Milky Way, from low-mass
and nearby clouds (Andr{\'e} et al. 2010) to most distant and
high-mass star-forming regions (Molinari et al. 2010).  As a
consequence, characterizing the physical properties of these filaments
has been revealed as key to our understanding of the origin of the
stars within molecular clouds.

The large majority of observational papers (Arzoumanian et al. 2011;
Palmeirim et al. 2013; Hacar et al. 2013) use the classical
``Ostriker'' profile (Ostriker 1964) as a benchmark to interpret
observations.  More specifically, if the estimated linear mass of an
observed filament is larger than the value obtained for the Ostriker
filament ($\simeq$ 16.6 M$_\odot$ pc$^{-1}$ for T=10 K), it is assumed
that the filament is unstable.  Analogously, density profiles flatter
than the Ostriker profile are generally interpreted as a a sign of
collapse.  However, it is worth recalling the assumptions and
limitations of this model: $(i)$ filaments are assumed to be
isothermal, $(ii)$ they are not rotating, $(iii)$ they are isolated,
$(iv)$ they can be modeled as cylindrical structures with infinite
length, $(v)$ their support against gravity comes solely from thermal
pressure.  An increasing number of observational results suggest
however that none of the above assumptions can be considered as
strictly valid.  In a first paper (Recchi et al. 2013, hereafter Paper
I) we have relaxed the hypothesis $(i)$ and we have considered
equilibrium structures of non-isothermal filaments.  Concerning
hypothesis $(ii)$, and after the pioneering work of Robe (1968), there
has been a number of publications devoted to the study of equilibrium
and stability of rotating filaments (see e.g. Hansen et al. 1976;
Inagaki \& Hachisu 1978; Robe 1979; Simon et al. 1981; Veugelen 1985;
Horedt 2004; Kaur et al. 2006; Oproiu \& Horedt 2008).  However, this
body of knowledge has not been recently used to constrain properties
of observed filaments in molecular clouds.  In this work we aim to
explore the effects of rotation on the interpretation of the physical
state of filaments during the formation of dense cores and stars.
Moreover, we emphasize the role of envelopes on the determination of
density profiles, an aspect often overlooked in the recent literature.

The paper is organised as follows. In Sect. \ref{sec:obs} we review
the observational evidences suggesting that star-forming filaments are
rotating.  In Sect.  \ref{sec:rotfil} we study the equilibrium
configuration of rotating filaments and the results of our
calculations are discussed and compared with available observations.
Finally, in Sect. \ref{sec:conc} some conclusions are drawn.

\section{Observational signs of rotation in filaments}
\label{sec:obs}

Since the first millimeter studies in nearby clouds it is well known
that star-forming filaments present complex motions both parallel and
perpendicular to their main axis (e.g. Loren 1989; Uchida et al.
1991).  Recently, Hacar \& Tafalla (2011) have shown that the internal
dynamical structure of the so-called velocity coherent filaments is
dominated by the presence of local motions, typically characterized by
velocity gradients of the order of 1.5--2.0 km~s$^{-1}$~pc$^{-1}$,
similar to those found inside dense cores (e.g. 
Caselli et al. 2002).  Comparing the structure of both density and
velocity perturbations along the main axis of different filaments,
Hacar \& Tafalla (2011) identified the periodicity of different
longitudinal modes as the streaming motions leading to the formation
of dense cores within these objects.  These authors also noticed the
presence of distinct and non-parallel components with similar
amplitudes than their longitudinal counterparts.  Interpreted as
rotational modes, these perpendicular motions would correspond to a
maximum angular frequency $\omega$ of about 6.5~$\cdot$ 10$^{-14}$
s$^{-1}$.  Assuming these values as characteristic defining the
rotational frequency in Galactic filaments, the detection of such
rotational levels then rises the question on whether they could
potentially influence the stability of these objects.\footnote{It is
  worth stressing that if the filament forms an angle $\beta \neq 0$
  with the plane of the sky, an observed radial velocity gradient
  $\frac{\Delta V_r}{\Delta r}$ corresponds to a real gradient that is
  $\frac{1}{\cos \beta}$ times larger than that.}

To estimate the dynamical relevance of rotation we can take the total
kinetic energy per unit length as equal to $\mathcal{T}=\frac{1}{2}
\omega^2 R_c^2 M_{lin}$, where $R_c$ is the external radius of the
cylinder and $M_{lin}$ its linear mass.  The total gravitational
energy per unit mass is $W=G{M_{lin}}^2$, hence the ratio
$\mathcal{T}/W$ is
\begin{equation} \frac{\mathcal{T}}{W} \simeq 0.65 \left(
    \frac{\omega}{6.5 \cdot 10^{-14}} \right)^2 \left(\frac{R_c}{0.15
      \,{\rm pc}} \right)^2 \left(\frac{M_{lin}}{16.6\, {\rm
        M}_{\odot} \,{\rm pc}^{-1}} \right)^{-1}.
  \end{equation}
\noindent 
Clearly, for nominal values of $\omega$, $R_c$ and $M_{lin}$ the total
kinetic energy associated to rotation is significant, thus rotation
is dynamically important.

\section{The equilibrium configuration of rotating, non-isothermal
filaments} \label{sec:rotfil}

In order to calculate the density distribution of rotating,
non-isothermal filaments, we extend the approach already used in Paper
I, which we shortly repeat here.  The starting equation is the
hydrostatic equilibrium with rotation: $\nabla P = \rho (g + \omega^2
r)$.  We introduce the normalization:
\begin{equation}
  \label{eq:normalization} 
  \rho = \theta \rho_0,\;\;T=\tau
  T_0,\;\; r=Hx\;\; \Omega=\sqrt{\frac{2}{\pi G \rho_0}} \omega.
\end{equation}
\noindent 
Here, $\rho_0$ and $T_0$ are the central density and temperature,
respectively, $H=\sqrt{\frac{2 k T_0}{\pi G \rho_0 \mu m_H}}$ is a
length scale and $\Omega$ is a normalized frequency.  Simple steps
transform the hydrostatic equilibrium equation into: 
\begin{equation}
  \theta\tau^\prime+\tau\theta^\prime=\theta\left(\Omega^2 x - 8
    \frac{\int_0^x {\tilde x} \theta d{\tilde x}}{x}\right).
  \label{eq:start}
\end{equation}

Calling now $I=\int_0^x {\tilde x} \theta d{\tilde x}$, then clearly
$I^\prime =\theta x$.  Solving the above equation for $I$, we obtain
$8I=\Omega^2
x^2 -\tau^\prime x - \tau x \frac{\theta^\prime}{\theta}$.  Upon
differentiating this expression with respect to $x$ and rearranging, we
obtain:
\begin{equation}
  \theta^{\prime\prime}=\frac{\left(\theta^\prime\right)^2}{\theta}
  -\theta^\prime\left[\frac{\tau^\prime}{\tau}+\frac{1}{x}\right]-
  \frac{\theta}{\tau}\left[\tau^{\prime\prime}+\frac{\tau^\prime}{x}+
  8 \theta -2 \Omega^2 -2 x \Omega\Omega^\prime\right].
\label{eq:basic} 
\end{equation} 
\noindent 
Correctly, for $\Omega=0$ we recover the equation already used in
Paper I.  This second-order differential equation, together with the
boundary conditions $\theta(0)=1$, $\theta^\prime(0)=-\tau^\prime(0)$
(see Paper I) can be integrated numerically to obtain equilibrium
configurations of both rotating and non-isothermal filaments
independently.  This expression is more convenient than classical
Lane-Emden type equations (see e.g. Robe 1968; Hansen et al. 1976) for
the problem at hand.  Notice also that also the normalization of
$\omega$ differs from the more conventional assumption
$\eta^2=\omega^2/4 \pi G \rho_0$ (Hansen et al. 1976).

\subsection{Uniformly rotating filaments}\label{subsec:rotfils}

\begin{figure} 
\begin{center} 
  \includegraphics[width=6cm, angle=270]{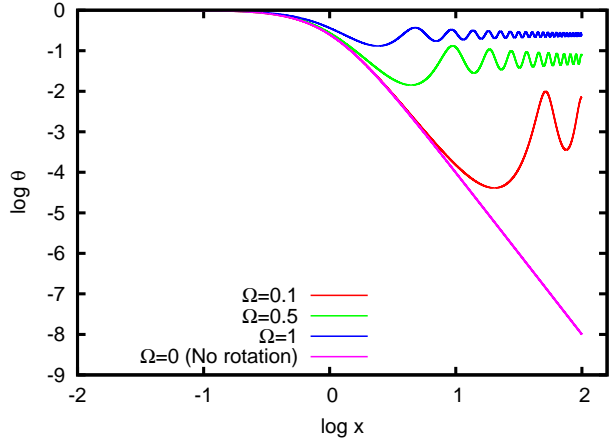}
  \caption{Logarithm of the normalized density $\theta$ as a function
    of $x$ for various models of isothermal filaments with different
    normalized angular frequencies.  The model with $\Omega=0$
    corresponds to the Ostriker profile with $\rho\propto r^{-4}$ at
    large radii.}
\label{fig:denrot} \end{center} \end{figure}
If we set $\tau, \Omega=Const$ in Eq. \ref{eq:basic}, we can obtain
equilibrium solutions for isothermal, uniformly rotating filaments. We
have checked that our numerical results reproduce the main features of
this kind of cylinders, already known in the literature, namely:
\begin{itemize}
\item Density inversions take place for $\Omega^2>0$ as the
  centrifugal, gravitational and pressure gradient forces battle to
  maintain mechanical equilibrium.  Density oscillations occur in
  other equilibrium distributions of polytropes (see Horedt 2004 for a
  very comprehensive overview).  Noticeably, the equilibrium solution
  of uniformly rotating cylindrical polytropes with polytropic index
  $n=1$ depends on the (oscillating) zeroth-order Bessel function
  $J_0$ (Robe 1968; see also Christodoulou \& Kazanas 2007).
  Solutions for rotating cylindrical polytropes with $n>1$ maintain
  this oscillating character although they can not be expressed
  analytically.  As evident in Fig. \ref{fig:denrot}, in the case of
  isothermal cylinders (corresponding to $n \rightarrow \infty$), the
  frequency of oscillations is zero for $\Omega=0$, corresponding to
  the Ostriker profile.  This frequency increases with the angular
  frequency $\Omega$.
\item For $\Omega>2$, $\rho^\prime(0)>0$, due to the fact that, in
  this case, the effective gravity $g+\omega^2r$ is directed outwards.
  For $\Omega<2$, $\rho^\prime(0)<0$.  If $\Omega=2$, there is perfect
  equilibrium between centrifugal and gravitational forces (Keplerian
  rotation) and the density is constant (see also Inagaki \& Hachisu
  1978).
\item The density tends asymptotically to the value $\Omega^2/4$.
  This implies also that the integrated mass per unit length $\Pi
  =\int_0^\infty 2 \pi x \theta(x) dx$ diverges for $\Omega^2>0$.
  Rotating filaments must be thus pressure truncated.  This limit of
  $\theta$ for large values of $x$ is essentially the reason why
  density oscillations arise for $\Omega \neq 2$.  This limit can not
  be reached smoothly, i.e.  the density gradient can not tend to
  zero.  If the density gradient tends to zero, so does the pressure
  gradient.  In this case there must be asymptotically a perfect
  equilibrium between gravity and centrifugal force (Keplerian
  rotation) but, as we have noticed above, this equilibrium is
  possible only if $\Omega=2$.  Thanks to the density oscillations,
  $\nabla P$ does not tend to zero and perfect Keplerian rotation is
  never attained.  Notice moreover that the divergence of the linear
  mass is a consequence of the fact that the centrifugal force
  diverges, too, for $x \rightarrow \infty$.
\end{itemize} 
All these features can be recognized in Fig.  \ref{fig:denrot}, where
the logarithm of the normalized density $\theta$ is plotted as a
function of the filament radius $x$ for models with various angular
frequencies $\Omega$, ranging from 0 (non-rotating Ostriker filament)
to 1.  Hansen et al. (1976) performed a stability analysis of
uniformly rotating isothermal cylinders, based on a standard linear
perturbation of the hydrodynamical equations.  They noticed that,
beyond the point where the first density inversion occurs, the system
behaves differently compared to the non-rotation case.  Dynamically
unstable oscillation modes appear and the cylinder tends to form
spiral structures.  Notice that a more extended stability analysis,
not limited to isothermal or uniformly rotating cylinders, has been
recently performed by Freundlich et al. (2014; see also Breysse et al.
2014).

Even in its simplest form, the inclusion of rotations has interesting
consequences in the interpretation of the physical state of filaments.
As discussed in Paper~I, the properties of the Ostriker filament
(Stod{\'o}lkiewicz 1963; Ostriker 1964), in particular its radial
profile and linear mass, are classically used to discern the stability
of these structures.  According to the Ostriker solution, an infinite
and isothermal filament in hydrostatic equilibrium presents an
internal density distribution that tends to $\rho (r) \propto r^{-4}$
at large radii and a linear mass
M$_{Ost}\simeq16.6$~M$_{\odot}$~pc$^{-1}$ at 10~K.  As shown in
Fig.~\ref{fig:denrot}, and ought to the effects of the centrifugal
force, the radial profile of an uniformly rotating filament in
equilibrium ($\Omega>0$) could present much shallower profiles than in
the Ostriker-like configuration (i.e.  $\Omega=0$).  Such departure
from the Ostriker profile is translated into a variation of the linear
mass that can be supported by these rotating systems. For comparison,
an estimation of the linear masses for different rotating filaments in
equilibrium truncated at a normalized radius x=3 and x=10 are
presented in Tables \ref{table1} and \ref{table2}, respectively. In
these tables, the temperature profile is the linear function
$\tau(x)=1+Ax$. In particular, the case $A=0$ refers to isothermal
filaments, whereas if $A>0$, the temperature is increasing
outwards.\footnote{In Paper I, we considered two types of temperature
  profiles as a function of the filament radius, i.e.
  $\tau_1(x)=1+Ax$ and $\tau_2(x)=[1+(1+B)x]/(1+x)$, whose constants
  defined their respective temperature gradients as functions of the
  normalized radius. Both cases are based on observations.  In this
  paper we will only consider the linear law $\tau=\tau_1(x)$; results
  obtained with the asymptotically constant law are qualitatively the
  same.} As can be seen there, the linear mass of a rotating filament
could easily exceed the critical linear mass of its Ostriker-like
counterpart without being necessary unstable.

It is also instructive to obtain estimations of the above models in
physical units in order to interpret observations in nearby clouds.
For typical filaments similar to those found in Taurus (Hacar \&
Tafalla 2011; Palmeirim et al. 2013; Hacar et al. 2013), with central
densities of $\sim 5\cdot 10^4$ cm$^{-3}$, one obtains $\Omega\simeq
0.5$ according to Eq.~\ref{eq:normalization}. Assuming a temperature
of 10~K, and from Tables \ref{table1} and \ref{table2} (case $A=0$),
this rotation level leads to an increase in the linear mass between
$\sim$~17.4 M$_\odot$~pc$^{-1}$ if the filament is truncated at radius
x=3, and up to $\sim$~112 M$_\odot$~pc$^{-1}$ for truncation radius of
x=10.  Here, it is worth noticing that a normalized frequency of
$\Omega\simeq 0.5$, or $\omega \sim 6.5 \cdot 10 ^{-14}$ s$^{-1}$,
corresponds to a rotation period of $\sim$~3.1~Myr.  With probably
less than one revolution in their entire lifetimes
($\tau\sim$~1--2~Myr), the centrifugal forces inside such slow
rotating filaments can then provide a non-negligible support against
their gravitational collapse, being able to sustain larger masses than
in the case of an isothermal and static Ostriker-like filament.
\subsection{Differentially rotating filaments} \label{subsec:diffrotfils}

\begin{figure} \begin{center} 
    \includegraphics[width=6cm, angle=270]{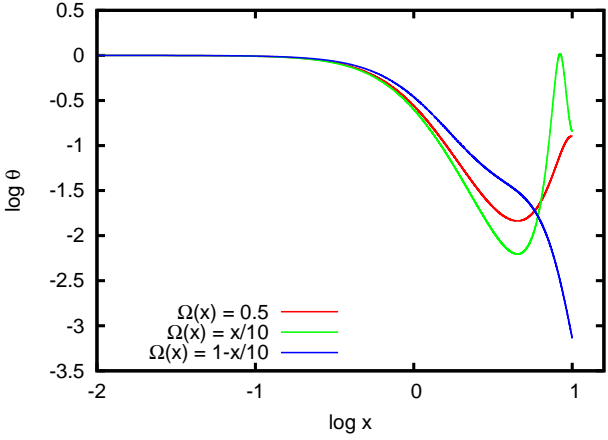}
    \caption{Logarithm of the normalized density $\theta$ as a
      function of $x$ for various models of filaments with different
      rotation laws.}
         \label{fig:dendiffrot}
\end{center} \end{figure}

As can be noticed in Fig.~\ref{fig:denrot}, a distinct signature of
the centrifugal forces acting within rotating filaments is the
presence of secondary peaks (i.e. density inversions) in their radial
density distribution at large radii.  Such density inversions could
dynamically detach the outer layers of the filament to its central
region, eventually leading to the mechanical breaking of these
structures.  In Sect.~\ref{subsec:rotfils}, we assumed that the
filaments present a uniform rotation, similar to solid bodies.
However, our limited information concerning the the rotation profiles
in real filaments invites to explore other rotation configurations.

\begin{table}
  \caption{Normalized linear masses at $x=3$ compared to the Ostriker
    filament with similar truncation radius, with M$_{Ost}(x\le 3)=
    14.9$~M$_{\odot}$~pc$^{-1}$, as a function of $\Omega$ and $A$.}
\label{table1} 
\centering 
\begin{tabular}{c c c c c} \hline\hline 
$\Omega$ & $A=0$& $A=0.02$ & $A=0.1$ & $A=0.5$\\ \hline 
0.1 & 1.006 & 1.015 & 1.049 & 1.167 \\ 
0.5 & 1.166 & 1.176 & 1.213 & 1.330 \\ 
0.8 & 1.553 & 1.561 & 1.593 & 1.676 \\ 
1.0 & 2.108 & 2.108 & 2.111 & 2.117 \\ \hline
\end{tabular} \end{table}

\begin{table}
  \caption{Similar to Table~\ref{table1} but for linear masses at
    $x=10$, with M$_{Ost}(x\le 10)= 16.4$~M$_{\odot}$~pc$^{-1}$.}
\label{table2} 
\centering 
\begin{tabular}{c c c c c} \hline\hline 
$\Omega$ & $A=0$& $A=0.02$ & $A=0.1$ & $A=0.5$\\ 
\hline 0.1 & 1.015 & 1.039 & 1.137 & 1.623 \\ 
0.2 & 1.075 & 1.102 & 1.212 & 1.730 \\ 
0.3 & 1.287 & 1.309 & 1.415 & 1.951 \\ 
0.4 & 2.533 & 2.321 & 2.063 & 2.379 \\ 
0.5 & 7.019 & 6.347 & 4.377 & 3.234 \\ 
0.6 & 10.37 & 10.53 & 9.398 & 4.988 \\ 
0.7 & 12.29 & 12.77 & 13.78 & 8.399 \\ 
0.8 & 14.96 & 15.14 & 16.59 & 13.84 \\ 
0.9 & 20.05 & 19.39 & 19.43 & 20.22 \\ 
1.0 & 26.22 & 25.70 & 23.71 & 25.95 \\ \hline
\end{tabular} \end{table}
  
For the sake of simplicity, we have investigated the equilibrium
configuration of filaments presenting differential rotation, assuming
that $\Omega$ linearly varies with the filament radius $x$.  For
illustrative purposes, we choose two simple laws: $\Omega_1(x)=x/10$
and $\Omega_2(x)=1-x/10$, both attaining the typical frequency
$\Omega=0.5$ at $x=5$. The first of these laws presumes that the
filament rotates faster at larger radii but presents no rotation at
the axis, resembling a shear motion.  Opposite to it, the second one
assumes that the filament presents its maximum angular speed at the
axis and that it radially decreases outwards.

The comparison of the resulting density profiles for these two models
presented above are shown in Fig. \ref{fig:dendiffrot} for normalized
radii x$\le$~10.  For comparison, there we also overplot the density
profile obtained with a constant frequency $\Omega=0.5$ (see
Sect.~\ref{subsec:rotfils}).  For these models, we are assuming A=0,
i.e. isothermal configurations.  Clearly, the law $\Omega_1(x)$
displays a radial profile with even stronger oscillations than the
model with uniform rotation.  As mentioned above, oscillations are
prone to dynamical instabilities.  In this case, instabilities start
occurring at the minimum of the density distribution, here located at
$x \simeq 4.45$.  Conversely, these density oscillations are
suppressed in rotating filaments that obey a law like $\Omega_2(x)$.
It is however worth noticing that this last rotational law fails to
satisfy the Solberg-H{\o}iland criterion for stability against
axisymmetric perturbations (Tassoul 1978; Endal \& Sofia 1978; Horedt
2004).  Stability can be discussed by evaluating the first order
derivative $\frac{d}{dx}[x^4 \Omega^2_2(x)]$, which is positive for $x
\in (0, 20/3) \cup (10, +\infty)$ and negative for $x \in (20/3, 10)$.
We must therefore either consider that this filament is unstable at
large radii, or we must assume it to be pressure-truncated at radii
smaller than x=20/3 $\simeq$~6.7.  As we mentioned above, we could not
exclude the hypothesis that rotation indeed induced instability and
fragmentation of the original filament, separating the central part
(at radii x$\simlt$ 4.45 for $\Omega=\Omega_1(x)$ and x$\simlt$ 6.7
for $\Omega=\Omega_2(x)$) from the outer mantel, which might
subsequently break into smaller units.  This (speculative) picture
would be consistent with the bundle of filaments observed in B213
(Hacar et al.  2013).  For comparison, the mass per unit length
attained by the model with $\Omega=\Omega_1(x)$ at $x<4.45$ (which
corresponds to $\sim$ 0.2 pc for $T=10$ K and $n_c \sim 5\cdot 10^4$
cm$^{-3}$) is equal to 0.99 M$_{Ost}$ whereas the mass outside this
minimum is equal to 22.7 M$_{Ost}$, i.e. there is enough mass to form
many other filaments.

\subsection{Non-isothermal and rotating filaments}\label{subsec:nonisofils}

\begin{figure} \begin{center} 
    \includegraphics[width=6cm, angle=270]{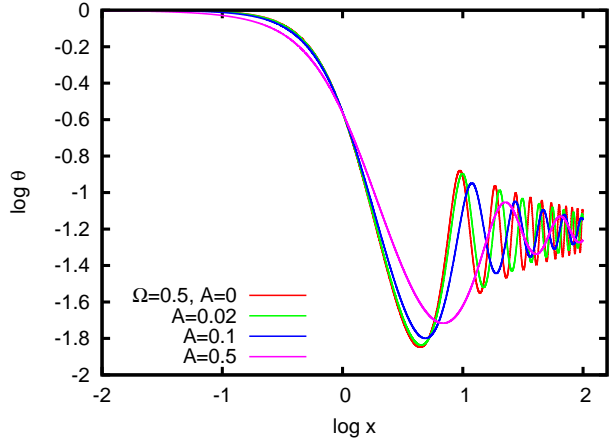}
    \caption{Logarithm of the normalized density $\theta$ as a
      function of $x$ for various models of uniformly rotating
      filaments with $\Omega=0.5$ and different temperature slopes
      $A$.}
\label{fig:denlinrot} \end{center} \end{figure}

As demonstrated in Paper I, the presence of internal temperature
gradients within filaments could offer an additional support against
gravity.  Under realistic conditions, these thermal effects should be
then considered in combination to different rotational modes in the
study of the stability of these objects.  

The numerical solutions obtained for the equilibrium configuration of
filaments with $\Omega=0.5$ and various values of $A$ are plotted in
Fig.  \ref{fig:denlinrot}.  Notice that Fig. 5 of Palmeirim et al.
(2013) suggests a rather shallower dust temperature gradient with a
value of A of the order of 0.02 (green curve in Fig.
\ref{fig:denlinrot}).  However, as discussed in Paper I, the gas
temperature profile could be steeper than the dust one, so it is
useful to consider also larger values of $A$.  Fig.
\ref{fig:denlinrot} shows that the asymptotic behaviour of the
solution does not depend on $A$: $\theta(x)$ always tends to
$\Omega^2/4$ for $x\rightarrow \infty$.  By looking at Eq.
\ref{eq:basic}, it is clear that the same asymptotic behaviour holds
for a wide range of reasonable temperature and frequency profiles.
Whenever $\tau^{\prime\prime}$, $\tau^\prime/x$ and
$\Omega\Omega^\prime x$ tend to zero for $x\rightarrow \infty$, and
this condition holds for a linear increasing $\tau(x)$ and for
$\Omega=$constant, the asymptotic value of $\theta(x)$ is
$\Omega^2/4$.  It is easy to see that also the asymptotically constant
law fulfils this condition if the angular frequency is constant.

Figure~\ref{fig:denlinrot} also shows that density oscillations are
damped in the presence of positive temperature gradients.  This was
expected as more pressure is provided to the external layers to
contrast the effect of the centrifugal force.  Since density
inversions are dynamically unstable, positive temperature gradients
must be thus seen as a stabilizing mechanism in filaments.  Our
numerical calculations indicate in addition that the inclusion of
temperature variations also increases the amount of mass that can be
supported in rotating filaments.  This effect is again quantified in
Tables \ref{table1} and \ref{table2} for truncation radii of $x=3$ and
$x=10$, respectively, compared to the linear mass obtained for an
Ostriker profile at the same radius.  As can be seen there, the
expected linear masses are always larger than in the isothermal and
non-rotating filaments, although the exact value depends on the
combination of $\Omega$ and $A$ due to the variation in the position
of the secondary density peaks compared to the truncation radius. 



\subsection{Derived column densities for non-isothermal, rotating
filaments: isolated vs. embedded configurations}

\begin{figure} \begin{center} 
    \includegraphics[width=6cm, angle=270]{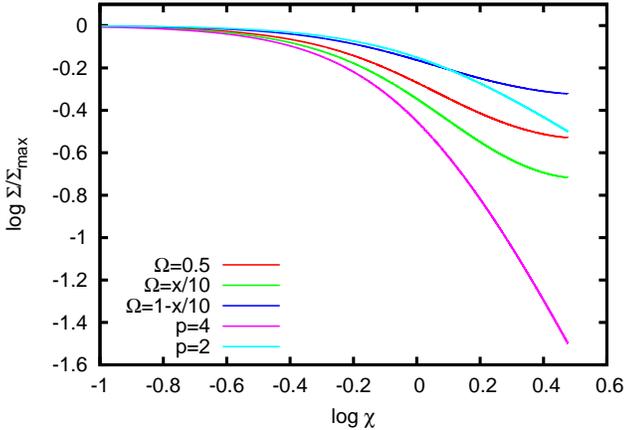}
    \caption{Column density, as a function of the normalized impact
      parameter $\chi$, for filaments characterized by three different
      rotation laws: increasing outwards ($\Omega=x/10$), decreasing
      outwards ($\Omega=1-x/10$) and constant ($\Omega=0.5$).  The
      filament is embedded in a cylindrical molecular cloud, with
      radius five times the radius of the filament.  The column
      density of the Ostriker filament (case $p=4$) and the one
      obtained for a Plummer-like model with $\rho\sim r^{-2}$ (case
      $p=2$) are also shown for comparison.}
\label{fig:dcolcyl} \end{center} \end{figure}

\begin{figure} \begin{center} 
    \includegraphics[width=6cm, angle=270]{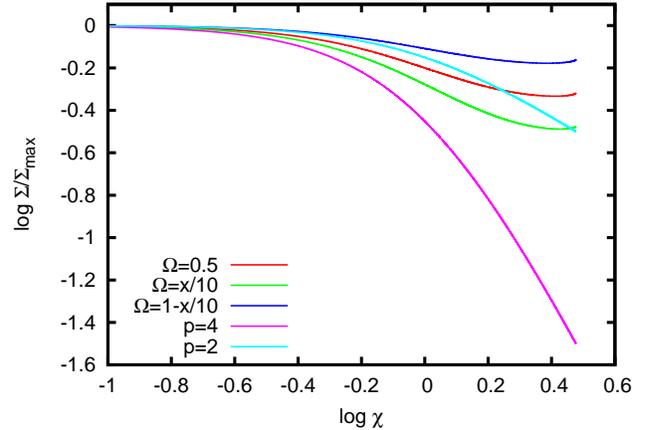}
    \caption{Same as Fig. \ref{fig:dcolcyl} but for a filament
      embedded in a slab, with half-thickness five times the radius of
      the filament.}
\label{fig:dcolslab} \end{center} \end{figure}

In addition to their radial profiles, we also calculated the column
density profiles produced by these non-isothermal, rotating filaments
in equilibrium presented in previous sections, as a critical parameter
to compare with the observations.  For the case of isolated filaments,
the total column density at different impact parameters $\chi$ can be
directly calculated integrating (either analytically or numerically)
their density profiles along the line of sight.  As a general rule, if
the volume density $\rho$ is proportional to $r^{-p}$, then the column
density $\Sigma (\chi)$ is proportional to $\chi^{1-p}$.  This result
holds not only for both Ostriker filaments (see also Appendix
\ref{sec:a2}) and more general Plummer-like profiles (e.g. see Eq.~1
in Arzoumanian et al. 2011), but also also for the new rotating,
non-isothermal configurations explored in this paper.  Recent
observations seem to indicate that those filaments typically found in
molecular clouds present column density profiles with $\Sigma(\chi)
\sim \chi^{-1}$, i.e. $p\simeq 2$ (see Arzoumanian et al. 2011;
Palmeirim et al. 2013), a value that we use for comparison hereafter.

An aspect often underestimated in the literature is the influence of
the filament envelope in the determination of column densities
profiles.  Particularly if a filament is embedded in (and
pressure-truncated by) a large molecular cloud, the line of sight also
intercepts some cloud material whose contribution to the column
density could be non-negligible (see also Appendix \ref{sec:a2}), as
previously suggested by different observational and theoretical
studies (e.g. Stepnik et al. 2003; Juvela et al. 2012).  In order to
quantify the influence of the ambient gas in the determination of the
column densities, here we consider two prototypical cases:
\begin{enumerate}
\item The filament is embedded in a co-axial cylindrical molecular
  cloud with radius $R_{m}$.
\item The filament is embedded in a sheet with half-thickness
  $R_{m}$.
\end{enumerate} Note that, if the filament is not located in the plane
of the sky, the quantity that enters the calculation of the column
density is not $R_{m}$ itself, but $R_{m}'=R_{m}/\cos\beta$, where
$\beta$ is the angle between the axis of the filament and this plane.

\begin{figure} \begin{center} 
    \includegraphics[width=6cm, angle=270]{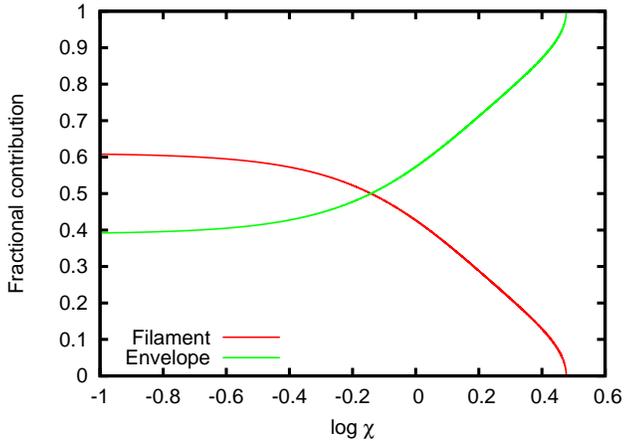}
    \caption{Fractional contribution of filament and envelope to the
      total column density.  The model shown here corresponds to the
      blue line of Fig. \ref{fig:dcolcyl}: the rotation profile is
      $\Omega=1-x/10$ and the filament is surrounded by a cylindrical
      envelope with R$_m$/R$_c$=5.}
\label{fig:fil_env} \end{center} \end{figure}

Following the results presented in
Sect.~\ref{subsec:rotfils}-\ref{subsec:nonisofils}, we have
investigated the observational properties of three representative
filaments in equilibrium obeying different rotational laws, namely
$\Omega_1(x)=x/10$, $\Omega_2(x)=1-x/10$ and $\Omega_3(x)=0.5$,
covering both differential and uniform rotational patterns.  The
contribution of the envelope to the observed column densities is
obviously determined by its relative depth compared to the truncation
radius of the filament as well as the shape of its envelope.  To
illustrate this behaviour, we have first assumed that these filaments
are pressure-truncated at $x=3$ (a conservative estimate).  Moreover,
we have considered these filaments to be embedded into the two
different cloud configurations presented before, that is a slab and a
cylinder, both with extensions $R_{m}$ corresponding to five times the
radius of the filament (i.e.  R$_{m}$/R$_{c}=5$).  In both cases, we
have assumed that the density of the envelope is constant and equal to
the filament density at its truncation radius i.e. at $x=3$.

The recovered column densities for the models presented above as a
function of the impact parameter $\chi$ in the case of the two
cylindrical and slab geometries are shown in Figs. \ref{fig:dcolcyl}
and \ref{fig:dcolslab}, respectively.  In both cases, the impact
parameter $\chi$ is measured in units of $H$.  The results obtained
there are compared with the expected column densities in the case of
two infinite filaments described by an Ostriker-like profile (case
$p=4$) and a Plummer-like profile with $\rho \propto r^{-2}$ at large
radii (case $p=2$), as suggested by observations.  From these
comparisons it is clear that all the explored configurations present
shallower profiles than the expected column density for its equivalent
Ostriker-like filament.  This is due to the constant value of the
density in the envelope, which tends to wash out the density gradient
present in the filament if the envelope radius is large.  Moreover,
the column densities expected for embedded filaments described by
rotating laws like $\Omega_1(x)$ and $\Omega_3(x)$ (this last one only
if the filament embedded into a slab) exhibit a radial dependency even
shallower than these p=2 models at large impact parameters.  The
relative contribution of filament and envelope is outlined in Fig.
\ref{fig:fil_env}.  The model shown here corresponds to the blue line
of Fig.  \ref{fig:dcolcyl}: the rotation profile is $\Omega=1-x/10$
and the filament is surrounded by a cylindrical envelope with
R$_m$/R$_c$=5.  As expected, at larger projected radii the observed
radial profiles are entirely determined by the total column density of
the cloud.

\begin{figure*} \begin{center} \vspace{-1cm} \includegraphics[width=11cm,
angle=270]{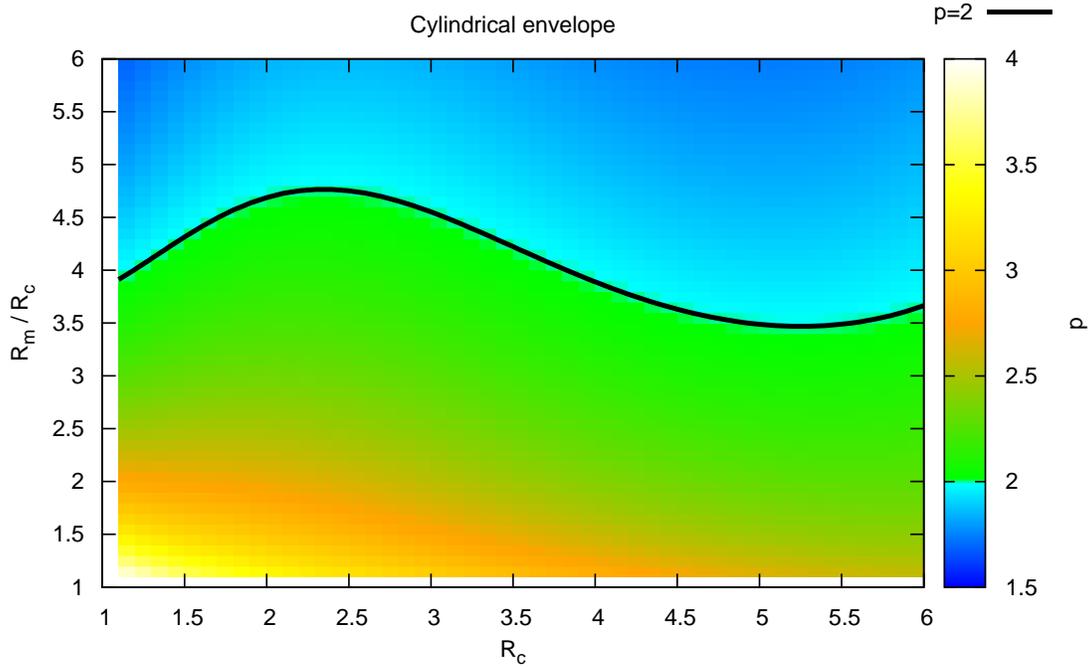} \vspace{-1cm}
\caption{Expected radial dependence for the observed column density
  profiles (colour coded) of rotating filaments in equilibrium obeying
  a rotation law like $\Omega_2(x)=1-x/10$, truncated at a radius
  R$_{c}$, and embedded into a cylindrical cloud extending up to a
  distance R$_{m}$.  R$_{c}$ and R$_{m}$ are displayed in units of the
  normalized (i.e. x) and truncation (i.e. R$_{m}$/R$_{c}$) radii,
  respectively. The black solid line highlights those models with a
  power-law dependence with p=2, similar to the observations.  Notice
  that also the color palette has been chosen in order to emphasize
  the transition from $p<2$ configurations to $p>2$ configurations.}
\label{fig:cyl_models} \end{center} \end{figure*}

Finally, it is important to remark that the expected column density
profiles for the models presented above and, particularly, their
agreement to these shallow Plummer-like profiles with p=2,
significantly depend on the selection of the truncation radius R$_{c}$
and the extent of the filament envelopes R$_{m}$.  This fact is
illustrated in Fig.~\ref{fig:cyl_models} exploring the expected slope
of the observed column density profiles for pressure truncated and
isothermal filaments following a rotational law like
$\Omega_2(x)=1-x/10$ under different configurations for both their
truncation and cloud radii. These results were calculated as the
averaged value of the local slope of the column density at impact
parameters $\chi\le R_{c}$, that is, where our models are sensitive to
the distinct contributions of both filaments and envelopes.  As
expected, the larger the cloud depth is compared to the filament, the
flatter profile is expected.  Within the range of values explored in
the figure, multiple combinations for both R$_{c}$ and R$_{m}$
parameters present slopes consistent to a power-law like dependency
with p=2.  Although less prominently, few additional combinations can
be also obtained in the case of filaments with rotational laws like
$\Omega_1(x)=x/10$ or $\Omega_3(x)=0.5$ (not shown here).  Unless the
rotational state of a filament is known and the contribution of the
cloud background is properly evaluated, such degeneration between the
parameters defining the cloud geometry and the relative weights of
both the filament and its envelope makes inconclusive any stability
analysis solely based on its mass radial distribution.

\section{Conclusions}
\label{sec:conc}

The results presented this paper have explored whether the inclusion
of different rotational patterns affect the stability of gaseous
filaments similar to those observed in nearby clouds.  Our numerical
results show that, even in configurations involving slow rotations,
the presence of centrifugal forces have a stabilizing effect,
effectively sustaining large amounts of gas against the gravitational
collapse of these objects.  These centrifugal forces promote however
the formation of density inversions that are dynamically unstable at
large radii, making the inner parts of these rotating filaments to
detach from their outermost layers.  To prevent the formation of these
instabilities as well as the asymptotical increase of their linear
masses at large radii, any equilibrium configuration for these
rotating filaments would require them to be pressure truncated at
relatively low radii.

In order to have a proper comparison with observations, we have also
computed the expected column density profiles for different pressure
truncated, rotating filaments in equilibrium.  To reproduce their
profiles under realistic conditions we have also considered these
filaments to be embedded in an homogeneous cloud with different
geometries.  According to our calculations, the predicted column
density profiles for such rotating filaments and their envelopes tend
to produce much shallower profiles than those expected for the case of
Ostriker-like filaments, resembling the results found in observations
of nearby clouds.  Unfortunately, we found that different combinations
of rotating configurations and envelopes could reproduce these
observed of profiles, complicating this comparison.  

To conclude, the stability of an observed filament can not be judged
by a simple comparison between observations and the predictions of the
Ostriker profile.  We have shown in this paper that density profiles
much flatter than the Ostriker profile and linear masses significantly
larger than the canonical value of $\simeq$ 16.6 M$_\odot$ pc$^{-1}$
can be obtained for rotating filaments in equilibrium, surrounded by
an envelope.  Detailed descriptions of the filament kinematics and
their rotational state, in addition to the analysis of their projected
column densities distributions, are therefore needed to evaluate the
stability and physical state in these objects.

\section*{Acknowledgements}
This publication is supported by the Austrian Science Fund (FWF). We
wish to thank the referee, Dr Chanda J. Jog, for the careful reading
of the paper and for the very useful report.

\appendix
\section{On the column density of filaments embedded in molecular
  clouds}
\label{sec:a2}
\begin{figure}
   \begin{center}
   \includegraphics[width=9cm]{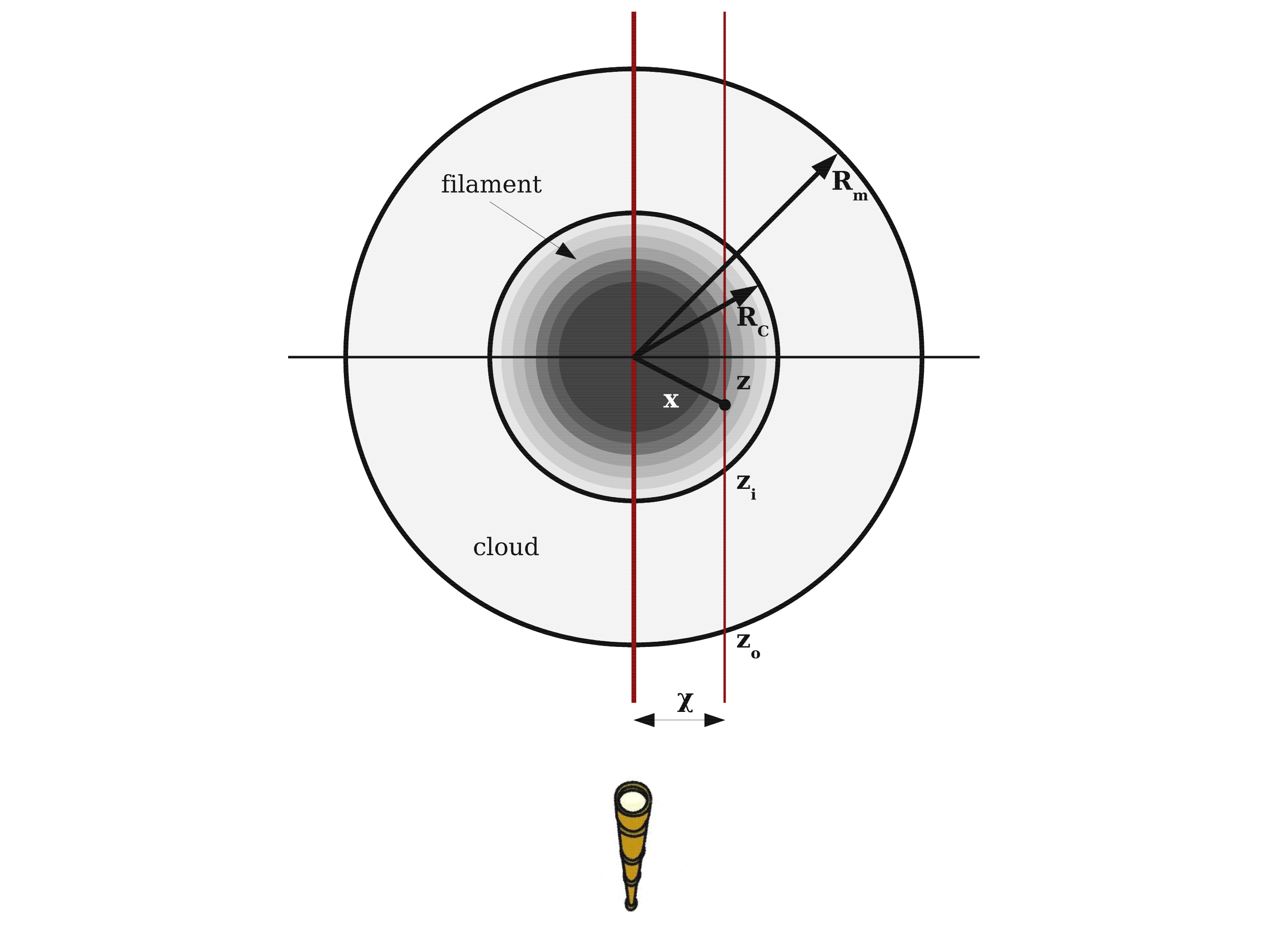}
   \caption{Section of the filament (with radius $R_c$), embedded in a
     (cylindrical, co-axial) molecular cloud with radius $R_m$.}
         \label{fig:cden_scheme}
   \end{center}
\end{figure}

In this appendix we derive a formula to calculate the column density
of filaments embedded in large molecular clouds.  For that, let us
assume first the general case of an isothermal filament described by
the Ostriker solution $\theta_i(x)=\left[1+x^2\right]^{-2}$.  If we
call $z$ the (normalized) distance between the plane in the sky where
the filament is located and a generic plane, then the distance between
the point $(\chi,z)$ (where $\chi$ is the normalized impact parameter)
and the axis is simply $x=\sqrt{\chi^2+z^2}$.  As it is well known, if
we assume that the filament extends until infinite distances, then the
column density is: 
\begin{align} 
\Sigma(\chi)&=\int_{-\infty}^\infty
  \theta_i(\chi,z)dz=\int_{-\infty}^\infty
  \frac{dz}{(1+z^2+\chi^2)^2}\notag\\
  &=\frac{1}{2}\frac{\pi}{(\chi^2+1)^{3/2}}. 
\end{align} 
\noindent
However, the cylinder could be embedded in a more extended cloud, with
radius $R_m$.  If we take for simplicity the cloud aligned with the
filament, the situation is shown in Fig.  \ref{fig:cden_scheme}.

Based on this figure (and due to the symmetry of the problem), we can
write the column density as:

\begin{equation} 
\Sigma(\chi)=2\int_{0}^{z_o} \theta_i(\chi,z)dz=
2\int_{z_o}^{z_i} \theta_b dz + 2 \int_0^{z_i}\frac{dz}{(1+z^2+\chi^2)^2}.
\end{equation} 
\noindent 
Here we have defined (see also Fig. \ref{fig:cden_scheme}): 
\begin{equation}
z_o=\sqrt{R_m^2-\chi^2},\;\;\;\;z_i=\sqrt{R_c^2-\chi^2},\;\;\;\;
\theta_b=\theta(R_c), 
\end{equation} 
\noindent 
and assumed that the density of the molecular cloud is constant and
equal to $\theta(R_c)$.  The result is:
\begin{align} \Sigma(\chi)&=2\theta_b
(z_o-z_i)+\frac{z_i}{(\chi^2+1)(R_c^2+1)} +\frac{\tan^{-1}
\sqrt{\frac{R_c^2-\chi^2}{\chi^2+1}}}{(\chi^2+1)^{3/2}}, \notag\\
&=2\frac{\sqrt{R_m^2-\chi^2}-\sqrt{R_c^2-\chi^2}}{(1+R_c^2)^2}+
\frac{\sqrt{R_c^2-\chi^2}}{(\chi^2+1)(R_c^2+1)} +\notag\\&
+\frac{\tan^{-1}
\sqrt{\frac{R_c^2-\chi^2}{\chi^2+1}}}{(\chi^2+1)^{3/2}}. \end{align}

\noindent 
It is easy to see that, in the limes for $R_c$ (and $R_m$) tending to
infinity, we recover the column density profile found above for the
infinite cylinder.

Another possibility is to assume that the cylinder is immersed in a
slab of gas with half thickness $R_m$.  The derivation of the column
density remains the same and the only difference is that $z_o$ is now
fix (it is equal to $R_m$) and does not depend any more on $\chi$ as
before.

For filaments whose profiles are determined numerically (like the ones
found in Sect. \ref{sec:rotfil}) the integral: \begin{equation}
  \int_0^{z_i}\theta(\chi,z)dz, \end{equation} \noindent (where as
usual $\chi$ and $z$ are related to $x$ by $x=\sqrt{\chi^2+z^2}$) must
be calculated numerically.  The contribution to the column density due
to the surrounding molecular cloud remains unaltered. 
\end{document}